\pdfoutput=1
\documentclass[fleqn,usenatbib,useAMS]{mnras}

\usepackage[T1]{fontenc}
\usepackage{ae,aecompl}

\usepackage[usenames,dvipsnames]{color}
\usepackage{rotating}
\usepackage{epstopdf}
\usepackage{color}
\usepackage[para,online,flushleft]{threeparttable}

\usepackage{natbib}
\bibpunct{(}{)}{;}{a}{}{,}

\usepackage{hyperref}

\usepackage{graphicx}	
\usepackage{amsmath}	
\usepackage{amssymb}	

\newcommand{\lsim}{\raise0.3ex\hbox{$<$}\kern-0.75em{\lower0.65ex\hbox{$\sim$}}}

\newcommand{\msun}{M$_{\odot}$}

\title[Satellites of Satellites]{Satellites of Satellites: The Case for Carina and Fornax}

\author[PARDY ET AL.]{
\parbox[t]{\textwidth}{Stephen A. Pardy$^{1}$,
Elena D'Onghia$^{1,2}$\thanks{Email: edonghia@astro.wisc.edu}, 
Julio Navarro$^{3}$, Robert Grand$^{4}$, Facundo A. G\'omez$^{5,6}$, Federico Marinacci$^{7}$, R\"udiger Pakmor$^{4}$, Christine Simpson$^{8}$, Volker Springel$^{4}$
}
\\ \\
\parbox[t]{\textwidth}{
$^1$Department of Astronomy, University of Wisconsin, 475 North Charter Street, Madison, WI 53706, USA\\
$^2$Center for Computational Astrophysics, Flatiron Institute, 162 Fifth Avenue, New York, NY 10010, USA\\
$^3$Senior CIfAR Fellow. Department of Physics and Astronomy, University of Victoria, Victoria, BC V8P 5C2, Canada\\
$^4$Max-Planck-Institut f\"{u}r Astrophysik, Karl-Schwarzschild-Str. 1, D-85748, Garching, Germany\\
$^5$Instituto de Investigaci\'on Multidisciplinar en Ciencia y Tecnolog\'ia, Universidad de La Serena, Ra\'ul Bitr\'an 1305, La Serena, Chile\\
$^6$Departamento de F\'isica y Astronom\'ia, Universidad de La Serena, Av. Juan Cisternas 1200 Norte, La Serena, Chile\\
$^7$Department of Physics \& Astronomy, University of Bologna, via Gobetti 93/2, 40129 Bologna, Italy\\
$^8$Department of Astronomy \& Astrophysics, University of Chicago, Chicago, IL 60637, USA \\
}
}

\date{Accepted XXX. Received YYY; in original form ZZZ}

\pubyear{2019}

\begin{document}

\label{firstpage}
\pagerange{\pageref{firstpage}--\pageref{lastpage}}
\maketitle

\begin{abstract} We use the Auriga cosmological simulations of Milky Way (MW)-mass galaxies and their surroundings to study the satellite populations of dwarf galaxies in $\Lambda$CDM. As expected from prior work, the number of satellites above a fixed stellar mass is a strong function of the mass of the primary dwarf. For galaxies as luminous as the Large Magellanic Cloud (LMC), and for halos as massive as expected for the LMC (determined by its rotation speed), the simulations predict about $\sim 3$ satellites with stellar masses exceeding $M_*>10^5\, M_\odot$. If the LMC is on its first pericentric passage, then these satellites should be near the LMC and should have orbital angular momenta roughly coincident with that of the LMC. We use 3D positions and velocities from the 2nd data release of the Gaia mission to revisit which of the ``classical'' MW dwarf spheroidals could plausibly be LMC satellites. The new proper motions of the Fornax and Carina dwarf spheroidals place them on orbits  closely aligned with the orbital plane of the Magellanic Clouds, hinting at a potential Magellanic association. Together with the Small Magellanic Cloud (SMC), this result raises to $3$ the number of LMC satellites with $M_*>10^5\, M_\odot$, as expected from simulations. This also fills the 12-mag luminosity gap between the SMC and the ultra-faints Hyi1, Car2, Hor1, and Car3, the few ultra-faint satellites confirmed to have orbits consistent with a Magellanic origin.
\end{abstract}


\section{Introduction}
\label{sec:intro}

In the current paradigm of structure formation, the Lambda-Cold Dark Matter scenario ($\Lambda$CDM), the mass function of substructures in a dark matter halo (``subhalos'') is approximately self-similar. This means that, expressed in units of the primary halo mass, the mass function of subhalos is  independent of the primary, and well approximated by a steep power-law \citep{BenMoore:1999ja,Springel:2008gd,BoylanKolchin:2009co,Wang:2012jg}.

The similarity is broken when considering the {\it stellar} mass function of satellite galaxies in clusters, groups, and around individual primaries. Clusters have far more ``substructure'' (i.e., satellite galaxies) than groups, and groups have more satellites than isolated bright primaries. The difference arises because the relation between galaxy stellar mass ($M_*$) and halo virial\footnote{We define the virial mass of a halo, $M_{200}$, as that enclosed within a radius, $r_{200}$, where the mean inner density is $200$ times the critical density for closure. We refer to virial quantities as those measured at or within that radius, and denote them with a ``200'' subscript.} mass, $M_{200}$, is highly non-linear, thus breaking the similarity.

Indeed, had $M_*$ and $M_{200}$ been related by a simple power law, then the scaled satellite luminosity function, $N(>\mu)$ (where $\mu=M_{*}^{\rm sat}/M_*^{\rm pri}$) would be independent of the primary mass. This independence is actually expected in the dwarf galaxy regime, where $\Lambda$CDM galaxy formation models (based on abundance-matching techniques and simulations) predict a steep, near power-law dependence of $M_*$ on $M_{200}$ over a wide range of stellar masses. In other words, the satellite luminosity function of dwarf primaries should be nearly independent of the stellar mass of the primary  \citep{Sales:2012el}.

In particular, these models predict roughly $\sim 4$-$5$ satellites within 10 magnitudes of the primary for dwarfs as luminous as the Large and Small Magellanic Clouds (hereafter LMC and SMC, respectively).  Because of its insensitivity to the actual primary mass, this is a robust prediction that may be tested by searching for dwarf galaxies that might have plausibly been associated with the Clouds at the time of their infall into the Milky Way (MW) halo.

The search for Magellanic satellites, in fact, has a long history, tracing back to the proposal by \citet{LyndenBell:1976fs} of a ``Greater Magellanic Galaxy'' system of dwarfs in the Milky Way halo \citep{LyndenBell:1982tk}. \citet{DOnghia:2008bu} emphasized that groups of dwarfs were a natural prediction of hierarchical models of galaxy formation, and went further to suggest that the LMC and SMC were the largest members of a group of dwarf galaxies that has been recently accreted into the Milky Way and that could include up to $7$ of the $11$ brightest MW satellites.

The dwarf group accretion idea was further elaborated using N-body simulations by a number of authors \citep{Li:2008dz,Lux:2010ht,Nichols:2011jp,Jethwa:2016vv,Deason:2016ar}, and has often been cited as a possible explanation for the dynamical peculiarities of the Milky Way satellite population \citep[see, e.g.,][for a recent review]{Pawlowski:2018gh}.

A major step toward identifying true Magellanic satellites came from new estimates of the proper motion of the LMC, made possible by painstaking astrometric measurements of Hubble Space Telescope images over different epochs \citep{Kallivayalil:2006uu}. These measurements showed that the tangential velocity of the LMC is much larger than its radial velocity, and also much larger than the expected circular velocity at its present distance of $\sim 50$ kpc from the Galactic centre \citep[see, e.g.,][for a recent review]{DOnghia:2015tc}. This implies long orbital periods and, in most currently favoured models, that the LMC is near its first pericentric approach to the Galaxy \citep[see, e.g.,][]{Besla:2007kg}.

If the LMC is on its first approach, then most Magellanic satellites should lie close to the LMC because the Galactic tidal field has not yet had enough time to disperse them \citep{Sales:2011bn}. Indeed, these authors identified the surroundings of the LMC as ``fertile hunting ground for faint, previously unnoticed MW satellites'', including Magellanic satellites as well. This 
prediction came spectacularly true with the recent discoveries of over 30 candidate dwarf galaxies in close sky proximity to the Clouds
thanks to the surveys like the Dark Energy Survey, the Magellanic Satellites Survey, SMASH and Pan-STARRS \citep{Koposov:2015wb, Bechtol:2015bd, DrlicaWagner:2015gb, Kim:2015ew, Kim:2015cq, Laevens:2015ep, Martin:2015tw, Luque:2016gf, DrlicaWagner:2016tb, Torrealba:2018, Koposov:2018tt}.

The first-approach scenario further restricts the range of distances, velocities, and sky positions of previously-associated Magellanic satellites, as they should follow the Clouds on orbits consistent with their tidal debris. Using these criteria, \citet{Sales:2011bn} found little evidence for a clear association between any of the ``classical'' (i.e., $M_V<-8$) dwarf spheroidals and the LMC: only the SMC emerged from that analysis as a clear companion of the LMC.

These authors further argued that the most stringent test of association with the Clouds is provided by the orbital angular momentum of a satellite (around the MW), whose direction must be roughly coincident with the Clouds. Using proper motion data available at the time, they ruled out Fornax and Carina as Magellanic satellites, despite reasonably favorable indications from their distances and radial velocities.  \citet{Sales:2017uz} and \citet{Kallivayalil:2018tx} extended this method to the ``ultra-faint'' (i.e., $M_V>-8$) satellite population, and their combined work was able to show that at least \emph{4} such satellites (Hor1, Car2, Car3, and Hyi1) are very likely associated with the Clouds.

Although this finding provides strong support for the hierarchical clustering of dwarfs, it also raises an interesting question about the luminosity function of the Magellanic association. The LMC has a fairly massive companion, the SMC, which is about $1.5$ mag fainter than the LMC, but its next most luminous satellite appears to be Hyi1, which is nearly 13 mag fainter. This leaves a $>10$ mag ``gap'' in the LMC satellite luminosity function that seems peculiar.

We revisit these issues here using cosmological hydrodynamical simulations of the formation of Milky Way-mass galaxies and their surroundings from the Auriga Project \citep{Grand:2017cd}. Our main aims are to explore the predictions of these models for the abundance of luminous satellites around galaxies like the LMC and to re-examine the association of some of the classical Milky Way dSphs with the Clouds using the latest proper motions and orbital parameters of MW satellites from ``Gaia DR2'', the second data release of the Gaia mission \citep[][]{Fritz:2018vu, Kallivayalil:2018tx, Simon:2018wv, Helmi:2018dl}.

This paper is organized as follows. In \autoref{sec:methods}) we discuss the Auriga simulations (\autoref{subsec:sim}) and how the simulated galaxy sample was selected and analyzed (\autoref{subsec:sample}). Our results are described in \autoref{sec:results} and a comparison with Gaia DR2 is presented in \autoref{sec:Gaia}. A summary of our main findings is given in \autoref{sec:conclusions}.

\begin{figure*}
   \includegraphics[width=7in]{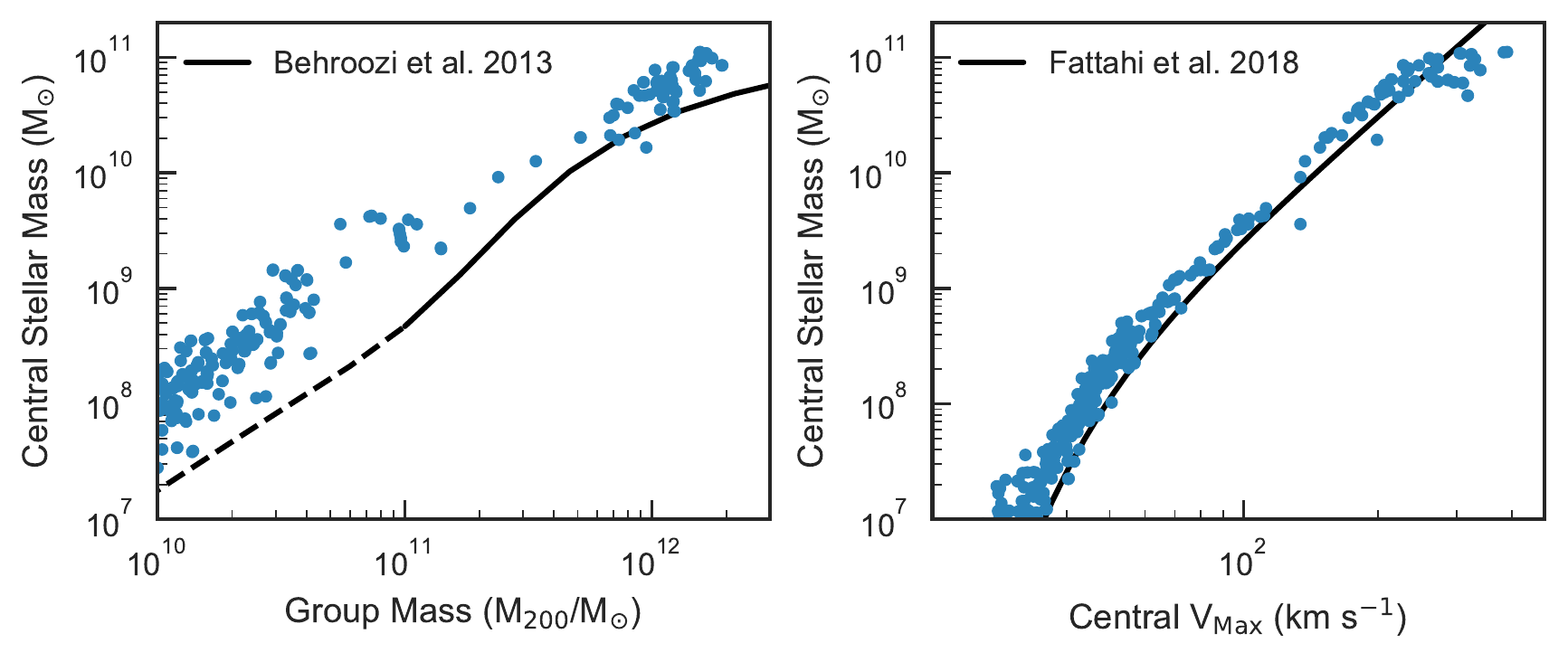} 
   \caption{\emph{Left:} Stellar mass versus virial mass for central galaxies in our Auriga sample (blue points), compared with the abundance-matching relation from \citet{Behroozi:2013fg} (black line). The dashed region of the \citet{Behroozi:2013fg} relation below 10$^{11}$ \msun\ is the extrapolation of their stellar mass-halo mass relation to lower masses. \emph{Right:} Galaxy stellar mass versus maximum circular velocity, V$_{\mathrm{max}}$, an alternative measure of halo mass, for central galaxies in our sample (blue points), compared with the equivalent relationship for the APOSTLE simulations reported by \citet{Fattahi:2017wd}. The central galaxy stellar mass is measured within r$_{\mathrm{gal}} = 0.15\, r_{200}$.}
   \label{fig:SMHM}
\end{figure*}

\section{Methods}
\label{sec:methods}

\subsection{The Auriga Simulations}
\label{subsec:sim}

We use 40 $\Lambda$CDM cosmological high-resolution magneto-hydrodynamic zoomed-in simulations of the formation of Milky Way analogues from the Auriga project \citep{Grand:2017cd}. The MW-analogues have halos with virial masses in the range between $5 \times 10^{11}$ and $2 \times 10^{12}\ M_{\odot}$.  The halos were identified at redshift $z = 0$ as isolated systems in a dark-matter-only simulation of a 100$^3$ Mpc$^3$ volume from the EAGLE project \citep{Schaye:2015gk}. The isolation criteria imply that each halo selected for resimulaton  is at least as far as  nine times the virial radius, $r_{200}$, of any other halo with mass greater than $3\%$ of the selected halo mass.

The initial conditions for the zoom re-simulations of the target halos were created at z = 127, and adopt the following cosmological parameters: $\Omega_m = 0.307$, $\Omega_b = 0.048$, $\Omega_{\Lambda} = 0.693$ and a Hubble constant of H$_0 = 100\, h $ km s$^{-1}$ Mpc$^{-1}$, where $h = 0.6777$. The selected halos are then re-simulated at higher resolution with full baryonic physics.

The simulations were performed with the moving-mesh code \textsc{AREPO}  \citep{Springel:2010hx}, including magneto-hydrodynamics \citep{Pakmor:2016jk} and a comprehensive galaxy formation model (see \citealt{Springel:2003eg, Vogelsberger:2013kw, Marinacci:2014il} and especially \citealt{Grand:2017cd}, for more details). The model includes: primordial and metal line cooling; a prescription for a uniform background UV field that completes reionization at z = 6 \citep{FaucherGiguere:2009ka}; a subgrid model for a multi-phase interstellar medium, star formation, and stellar feedback; black hole seeding, accretion, and feedback; and magnetic fields.  

In this paper, we focus on the medium-resolution simulations of the AURIGA suite, which correspond to the ``level 4'' resolution described in \citet{Grand:2017cd}. The typical dark-matter particle mass is $\sim3\times 10^5\, M_\odot$, and the baryonic mass resolution is $\sim5\times 10^4\, M_\odot$. The physical softening of collisionless particles is fixed in comoving coordinates and increases in physical units up to a maximum length of $369$ pc, which is reached at $z=1$. The physical-softening value for the gas cells is scaled by the gas-cell radius (assuming a spherical cell shape given the volume), with a minimum gravitational softening set to that of the collisionless particles.

Each high-resolution volume is embedded in a larger region containing progressively higher-mass boundary particles. The uncontaminated high resolution region typically extends beyond $3\times r_{200}$, but is highly non-spherical.

\subsection{Simulated galaxy sample selection}
\label{subsec:sample}

Structures in the Auriga volumes are identified using a two-step process. First the standard ``friends-of-friends (FOF)'' algorithm \citep{1985ApJ...292..371D} groups particles together using a fixed linking length, chosen to be 0.2 of the mean interparticle separation. Then the ``subfind'' algorithm \citep{Springel:2001cp} recursively identifies gravitationally bound groups, identifying the subhalos within each FOF halo.

For this work, we select all unique FOF groups that are uncontaminated by boundary particles at the present day (they must be more than 750 kpc from the nearest boundary particle). The central galaxy of each FOF group is designated as the ``primary'' or ``host'' galaxy of its group, and all of its subhalos within two virial radii of each group are labelled as its ``satellites''. Note that many low-mass subhalos contain no stars; i.e., they are ``dark''. We measure the stellar mass and other quantities of interest of central host galaxies using only particles within $r_{\rm gal}=0.15\, r_{200}$.

\section{Results}
\label{sec:results}

\autoref{fig:SMHM} shows the stellar mass vs halo virial mass (SMHM) relation for all primary galaxies in our sample, selected at $z=0$. This figure also shows the dependence of the stellar mass on the maximum circular velocity of each system, $V_{\rm Max}$, an alternative measure of halo mass that is more appropriate for satellites, for which a virial mass cannot be defined.

For reference, we show the abundance-matching relation of \citet{Behroozi:2013fg} in the left-hand panel of \autoref{fig:SMHM}. Auriga galaxies seem to have, on average,  more stars than indicated by the abundance-matching relation, a result discussed in detail by \citet{Grand:2017cd} and \citet{Simpson:2017tj}. We also include the $M_*$-$V_{\rm Max}$ relation reported by \citet{Fattahi:2017wd} for the APOSTLE suite of Local Group cosmological simulations run with the EAGLE code. These have been shown to match fairly well the mass function and internal structure of Local Group dwarfs \citep{Sawala:2016}. 

Auriga again seems to predict slightly more massive galaxies than APOSTLE over the whole halo mass range. For the purposes of this analysis, this may be taken to suggest that our predictions for the satellite luminosity function of dwarfs should be treated as upper limits.

Note as well that both the SMHM and $M_*$-$V_{\rm Max}$ relation are rather tight. This implies that for given stellar mass, the halo mass of a primary galaxy is well constrained. The reverse is not true, especially at low masses, where the steepness of the relation precludes an accurate prediction of the stellar mass of a galaxy at given halo mass.

\begin{figure*}
   \centering
   \includegraphics[width=7in]{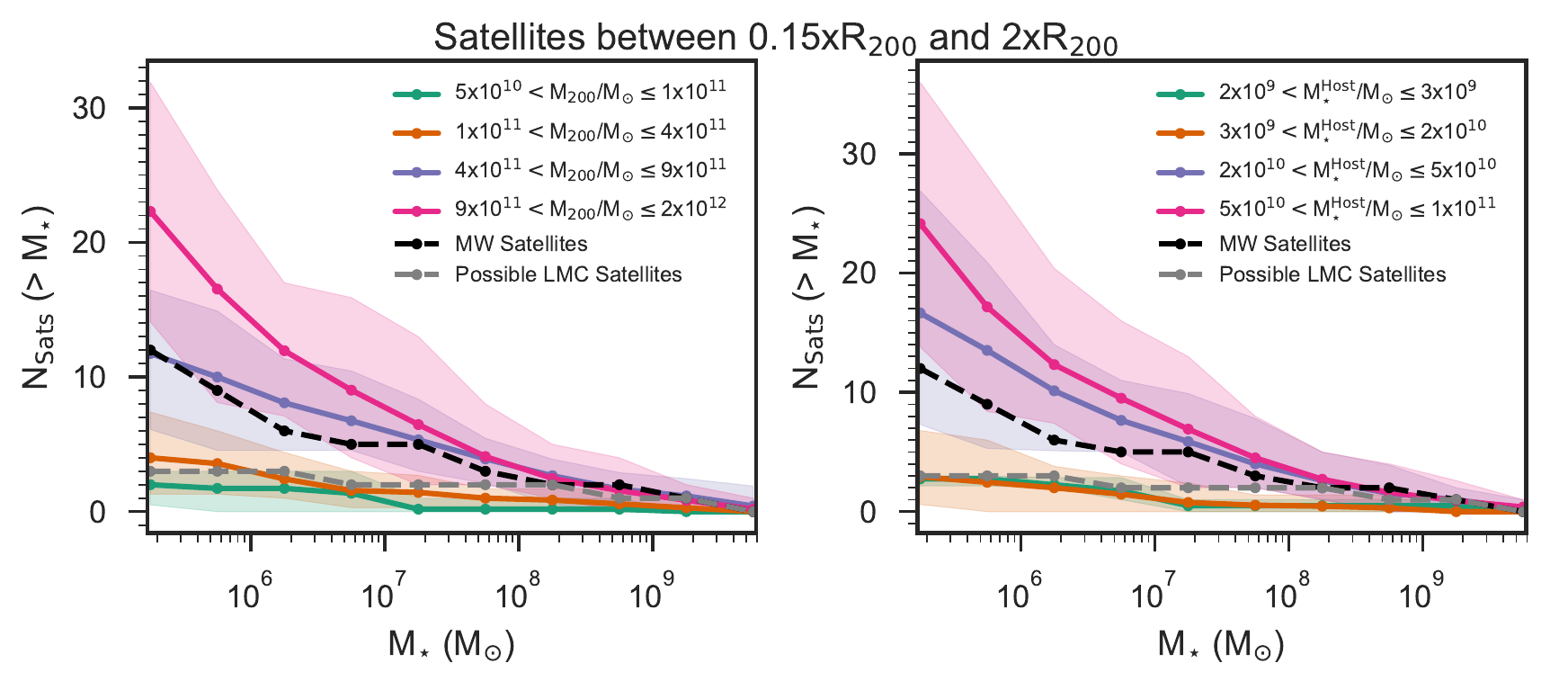} 
   \caption{ The cumulative number of satellites above a given stellar mass.  We measure this quantity for FOF groups selected by mass, and show the median (solid line with points) and inner 90\% of the distribution (shaded region) for the cumulative abundance function of all FOF groups in each selection. We also include the observed cumulative satellite stellar mass function for the Milky Way from \citet{McConnachie:2012fha}, which we show in the black dashed line. We also show the possible LMC satellites: SMC, Fornax, Carina (in order of decreasing mass) as a gray dashed line. \emph{Left:} FOF groups selected by M$_{200}$. \emph{Right:} FOF groups selected by central galaxy M$_{\star}$.}
   \label{fig:mstarDistribution}
\end{figure*}

\subsection{Satellites of dwarfs}

We explore next the satellite population of the primary galaxies shown in \autoref{fig:SMHM}. Although the subhalo mass function is expected to be self-similar, as discussed in \autoref{sec:intro}, the luminous satellite population is expected to depend strongly on halo mass and, consequently, on the stellar mass of the host galaxy. We show this in \autoref{fig:mstarDistribution}, where we have grouped the centrals in bins of halo mass (left-hand panel) and of central stellar mass (right-hand panel). The number of ``luminous'' satellites, defined as those with $M_{\star}>10^5\, M_\odot$ (roughly the stellar mass of the Draco dwarf spheroidal, the faintest of the ``classical'' dSphs with $M_V\sim -8$), is clearly a strong function of the mass of the system. For an LMC-like primary, with a stellar mass of $2 \times 10^9\, M_\odot$, these results indicate that we should expect of order $\sim 3$ satellites at least as bright as Draco. We should caution the reader that satellites with masses of $\sim M_{\star}>10^5\, M_\odot$ could contain as few as two stellar particles and the mass discretization may affect our results.

One could also use the halo mass to estimate the expected number of satellites, but this requires assuming a total halo mass for the LMC. Abundance-matching suggests a virial mass for the LMC of $1.6\times 10^{11}\, M_\odot$ \citep{Moster:2010ep, 2011MNRAS.413..101G}. Alternatively, assuming that the rotation speed of the LMC \citep[about $70$ km/s according to][]{Alves:2000hl} matches the virial velocity of its halo indicates a total mass of order $\sim 1.1 \times 10^{11}\, M_\odot$. The study of \citet{vanderMarel:2014bi} reports a higher circular velocity of $\sim 90$ km/s within $\sim 9$ kpc, implying an even higher virial mass of order $\sim 2.4\times 10^{11}\, M_\odot$. Examining the presence of satellites in Auriga as massive as the SMC ($M_{\star} > 10^9$ \msun) further supports a heavy LMC. We find no satellites as massive as the SMC around lower-mass hosts (those with M$_{200}$ $<$ 10$^{11}$). Although rare, we do find SMC-massed satellites around higher-mass hosts (those with M$_{200}$ between 1$\times$10$^{11}$ and 4$\times$10$^{11}$). Future kinematic observations in the halo may reveal the presence of a wake from the interaction between the LMC and the MW which would provide a strong constraint for the total mass of the LMC \citep{GaravitoCamargo:2019wg}. In any case, the predicted number of ``luminous'' satellites is still expected to be of order $\sim 3$ or more. 

The LMC has one known satellite, the SMC, but the next brightest confirmed satellite, Hyi1, has $M_V \sim -4.7$, or a stellar mass of order $\sim 5\times 10^3 M_\odot$ \citep{Kallivayalil:2018tx}. Does this suggest that the halo mass of the LMC is substantially smaller than suggested by either abundance-matching or its rotation speed? Or that somehow some of the classical dSphs have a Magellanic origin, as suggested by \citet{LyndenBell:1976fs} and \citet{DOnghia:2008bu}?

\subsection{The Magellanic association of classical dSphs}
\label{sec:Gaia}

The possible association of the classical MW dSphs with the Clouds was studied in detail by \citet{Sales:2011bn}, who concluded that there was no strong evidence for any of them to have a Magellanic origin. However, these authors also cautioned that their conclusion should be revisited when better proper motion data became available. Their criteria for Magellanic association hinges on a few simple indicators: (i) sky proximity to the Clouds, (ii) position, distance, and radial velocity consistent with tidal debris from the Clouds; and (iii) orbital angular momentum direction coincident with the Clouds'.

Criterion (i) is motivated by the fact that, if the Clouds are on their first pericentric passage about the Milky Way then the tidal field has not yet had time to fully disperse the Magellanic group. Most such satellites, like the Clouds, are expected to be near the pericenter of their orbits around the Milky Way.

Criterion (ii) applies to satellites that lag behind or have sped ahead of the Clouds, following approximately the orbital plane traced by the Magellanic orbit. If behind the Clouds, the satellite must still be infalling, and therefore should have negative Galactocentric radial velocity, while if ahead of the Clouds the satellite must be past pericenter, and should therefore have positive radial velocities.

Finally, criterion (iii) can only be applied to systems with accurate 3D velocity and position estimates, but is perhaps the most telling. This is because  the Magellanic system is much less massive than the Milky Way and its orbital velocity today is much higher than the likely velocity dispersion of its satellite system. Therefore, all of its associated satellites must approximately share the same orbital plane, which  implies that the direction of their orbital angular momenta must be very similar to that of the LMC.

Following these criteria, the Sagittarius dSph is easily excluded because its orbital plane is nearly perpendicular to that of the LMC (i.e., the Magellanic Stream is nearly perpendicular to the Sagittarius stream, although they both roughly trace polar orbits). Draco and Ursa Minor are so far away from the Clouds (almost diametrically opposite on the sky) that it is hard to make a case for association. Sextans and Sculptor have radial velocities inconsistent with their position on the sky if they had been stripped from the LMC. Tucana and Hercules are also easily excluded given their distances, positions, and velocities. We refer the interested reader to \citet{Sales:2011bn} for details.

On the other hand, the SMC passed all of these criteria for association. Fornax and Carina were problematic. Although they passed criteria (i) and (ii), the proper motions available at the time indicated that their orbits were not aligned with that of the Clouds. It is important then to review the possible association of these systems now that more precise proper motions are available from the Gaia DR2.

\begin{figure*}
   \centering
   \includegraphics[width=7in]{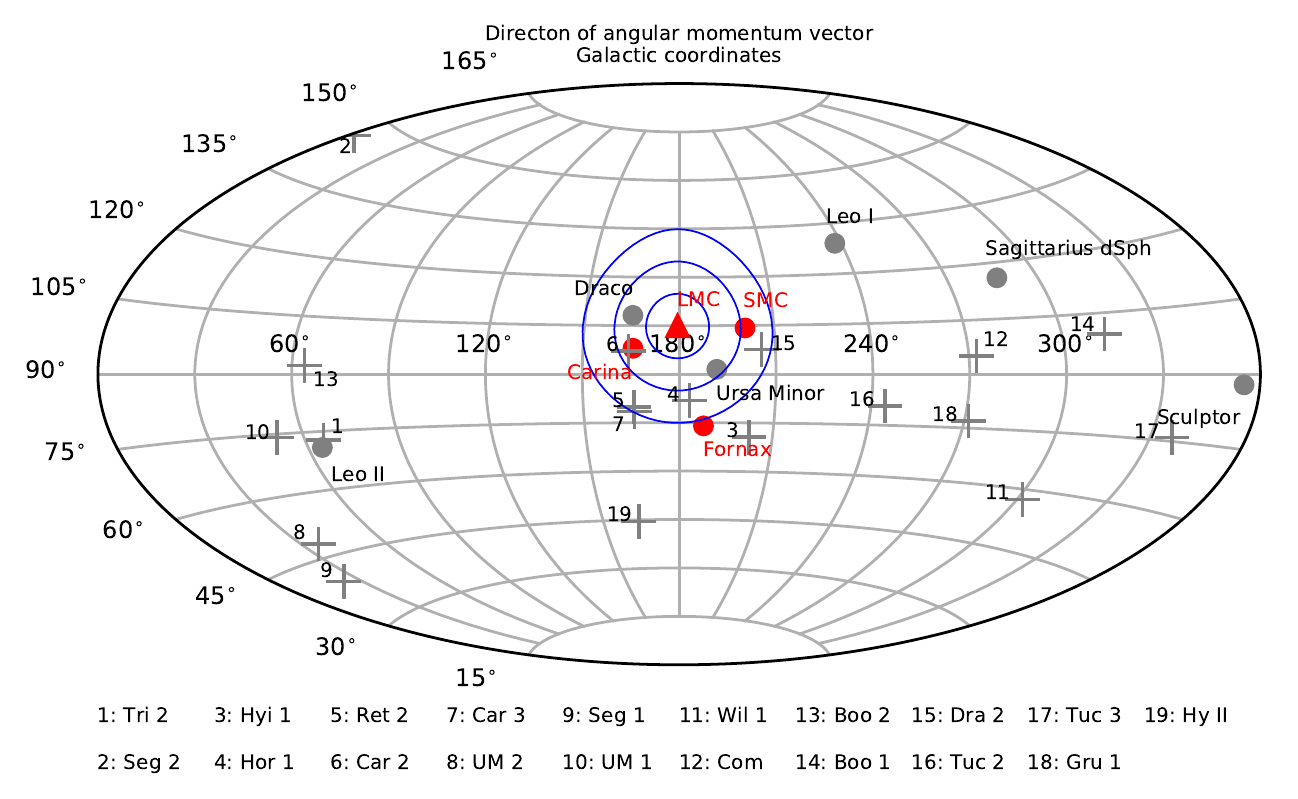} 
   \caption{The direction of the orbital angular momentum vectors in Galactocentric coordinates for Milky Way dwarf galaxies, using data from \citet{McConnachie:2012fha, Sales:2017uz, Fritz:2018vu, Helmi:2018dl, Simon:2018wv}. Luminous dwarf galaxies (i.e., $M_V<-8$)  are shown as circles, while ultrafaint galaxies are shown as crosses. Luminous dwarfs that are possible candidate of the LMC are shown in red, others are shown in gray. The blue circles which have radii of 10$^{\circ}$, 20$^{\circ}$, and 30$^{\circ}$ from the direction of the LMC's angular momentum.}
   \label{fig:angularmomentum}
\end{figure*}

Using velocities as reported by  \citet{Helmi:2018dl}, and auxiliary data from \citet{McConnachie:2012fha, Sales:2017uz, Fritz:2018vu, Simon:2018wv}, we compute the orbital angular momentum of all $10$ ``classical'' MW dSphs and many of the ultrafaint dwarf galaxies. We show the direction of the angular momentum vector projected onto galactic coordinates in \autoref{fig:angularmomentum}. We indicate in red the $3$ satellites that were selected based on the location and angular momentum criteria above. The contours indicate loci of fixed $\alpha$, where $\alpha$ is the angle between $\vec{L}_{\rm LMC}$ and $\vec{L}_{\rm sat}$. The new measurements indicate that Fornax and Carina, which \citet{Sales:2011bn} had concluded were unlikely Magellanic companions, are now clear candidates for Magellanic association.

The addition of Fornax and Carina would bring to $3$ the total number of dwarfs more massive than $10^5\, M_\odot$ associated with the LMC and would  also fill the ``luminosity gap'' in the Magellanic satellite luminosity function between the SMC and Hyi1, bringing it into excellent agreement with the results of the simulations discussed above.

Before these conclusions can be fully accepted, however, a number of issues need to be resolved. One is that, at least when using the MW mass model adopted by \citet{Helmi:2018dl}, the eccentricity of Fornax's orbit is quite different from that of the LMC. Whereas the LMC is on a highly radially biased orbit, Fornax appears to follow an orbit much closer to circular. Another is that although Fornax is close in the sky to the Clouds, it is actually much farther away, at a distance of $\sim 180$ kpc, compared to $\sim 50$ kpc for the LMC.

Addressing these issues satisfactorily requires more sophisticated modeling, where the accretion of the Magellanic system in the evolving Galactic potential is taken into account. For example, Fornax orbits are usually computed assuming a static spherical halo potential, which is bound to be a poor approximation, especially in the southern hemisphere and near the Clouds, whose own halo may disturb the orbit calculations and deflect the motions of other satellites \citep[see, e.g.,][]{Erkal:2018}. These questions are best addressed with direct numerical simulations, where a realistic population of Magellanic satellites should be evolved in a live, evolving MW+LMC potential. We plan to address these issues in future work.

\section{Conclusions}
\label{sec:conclusions}

We have studied the mass function of the satellites of dwarf galaxies using the Auriga project, a set of 40 high-resolution magneto-hydrodynamical simulations of the formation of Milky Way-like galaxies in the $\Lambda$CDM cosmogony. These simulations indicate that most isolated galaxies as massive as the Large Magellanic Cloud should be accompanied, on average, by about $\sim 3$ satellites at least as massive as $M_{\star}=10^5\, M_\odot$, or, equivalently, at least as luminous as $M_V=-8$. This theoretical expectation is at odds with the results of earlier work, who argued that, aside from the Small Magellanic Cloud, the second brightest confirmed Magellanic satellite would be Hyi1, with an absolute magnitude of $-4.7$ \citep{Sales:2011bn,Sales:2017uz,Kallivayalil:2018tx}. This implies an unexpected deficit of relatively luminous Magellanic satellites and a surprising gap of more than 10 magnitudes in its satellite luminosity function.

This result may be explained in a number of ways. One is simply to accept that the Clouds have an odd assortment of satellites, with an overly massive one (the SMC) and a large gap to the ultra-faint regime. Another is that some bright satellites may have been missed because of their extreme low surface brightness. The example of Crater II, a fairly massive ($M_*\sim 10^5\, M_\odot$) MW satellite only discovered in 2016 because of its unusually low surface brightness \citep{Torrealba:2016hq}, gives credence to this possibility. Finally, there is the possibility that some of the classical MW dSphs are actually Magellanic satellites, as suggested by \citet{DOnghia:2008bu}.

We have revisited the possibility that the Fornax and Carina dSphs might be associated with the Clouds using new proper motions from Gaia DR2 and find that these make them strong candidates for being Magellanic satellites. Indeed, the new 3D velocities of these satellites put them on orbits with angular momentum directions closely aligned with that of the Clouds. The Draco and Ursa Minor dSphs also share this orbital plane, but their position in the northern sky, almost diametrically opposite to the Clouds, lessens the likelihood of association. The proximity of Fornax and Carina to the Clouds in the southern sky bolsters the argument for association.

The addition of Fornax and Carina to the Magellanic trove of satellites would resolve the puzzling gap in its satellite luminosity function, bringing observations in close agreement with simulation expectations. Further work should focus on whether this association can be disproved using fully self-consistent, live direct numerical simulation of the evolution of Magellanic analogues, including its satellites, as they are accreted into the evolving Galactic potential.

\section{Acknowledgements}

E.D.O. acknowledges support from the Vilas Associate Research Fellowship and thanks the center for Computational Astrophysics for the hospitality during the completion of this work. JFN acknowledges the hospitality of the KITP at UC Santa Barbara and of the Aspen Center for Physics, where this research was partially carried out.  FM acknowledges support through the Program ``Rita Levi Montalcini'' of the italian MIUR. KITP is supported in part by the National Science Foundation under Grant No. NSF PHY-1748958. The ACP is supported by National Science Foundation grant PHY-1607611. This research made extensive use of many open-source Python packages including the Pathos\thanks{http://trac.mystic.cacr.caltech.edu/project/pathos} multiprocessing library \citep{McKerns:2012wd}, and the Scipy ecosystem \citep{Perez:hy, Hunter:ih, vanderWalt:dp}.

\bibliographystyle{mnras}
\bibliography{bibtex_library}
\label{lastpage}
\end{document}